\documentclass[twocolumn,notitlepage,pra,superscriptaddress]{revtex4-2}
\usepackage{graphicx}
\usepackage{braket}
\usepackage{amsmath,amsfonts,amssymb,amstext,amscd,amsthm,bbold}
\usepackage{comment,float}
\usepackage{tabularx}

\usepackage[colorlinks=true,linkcolor=blue,urlcolor=magenta,citecolor=magenta]{hyperref}
\usepackage[normalem]{ulem}
\usepackage{dsfont}
\usepackage{braket}
\usepackage{rotating}
\usepackage{subfig}
\usepackage{lipsum}
\usepackage{CJK}
\usepackage{multirow}
\usepackage{lipsum}

\begin{document}
	

	\title{Quantum illumination using polarization-entangled photon pairs for enhanced object detection}
	
	
	\author{Kanad Sengupta }
	\affiliation{Dept of Instrumentation and Applied Physics, Indian Institute Science, Bengaluru 560012, India}
	\author{K.Muhammed Shafi}
	\affiliation{Dept of Instrumentation and Applied Physics, Indian Institute Science, Bengaluru 560012, India}
	\author{Soumya Asokan}
	\affiliation{Dept of Instrumentation and Applied Physics, Indian Institute Science, Bengaluru 560012, India}
	\author{C. M. Chandrashekar}
	\affiliation{Dept of Instrumentation and Applied Physics, Indian Institute Science, Bengaluru 560012, India}
	\affiliation{The Institute of Mathematical Sciences, C. I. T. Campus, Taramani, Chennai 600113, India}
	\affiliation{Homi Bhabha National Institute, Training School Complex, Anushakti Nagar, Mumbai 400094, India}
	
	\begin{abstract}
		Entangled light sources for illuminating objects offer advantages over conventional illumination methods by enhancing the detection sensitivity of reflecting objects. The core of the quantum advantage lies in effectively exploiting quantum correlations to isolate noise and detect objects with low reflectivity. This work experimentally demonstrates the benefits of using polarization-entangled photon pairs for quantum illumination and shows that the quantum correlation measure, using CHSH value and normalized CHSH value, is robust against losses, noise, and depolarization. We report the detection of objects with reflectivity ($\eta$) as low as 0.05 and an object submerged in noise with a signal-to-noise ratio of 0.003 using quantum correlation and residual quantum correlation measures, surpassing previous results. Additionally, we demonstrate that the normalized CHSH value aids in estimating the reflectivity of the detected object. Furthermore, we analyze the robustness of the correlation measure under photon attenuation in atmospheric conditions to show the practical feasibility of real-time applications. 
	\end{abstract}
	
	\maketitle
	
	\section{Introduction}

	Quantum correlations, notably exemplified by entanglement exhibits genuine quantum character in composite systems and forms the intriguing resource for practical applications in quantum information processing protocols\,\cite{FSS19, PHF16, JSW2000, BPM97, PBG18}. However, their vulnerability to environmental disturbances poses a significant challenge, potentially undermining the advantages derived from such nonclassical relationships. An intriguing exception to this sensitivity is found in the realm of quantum illumination (QI)\,\cite{SL08,TEG08GS, LBD13, UR09, YB2020, HK2020, DGE2019, MLS17, QIReview2023, QIReview2023_2, BGW14}, an entanglement-based protocol for low reflectivity object detection under noisy thermal background. In the last decade, there have been significant efforts and developments to extend the advantages reported in QI schemes to potential areas, including quantum sensing\,\cite{LUR2017}, quantum bio imaging\,\cite{TRU2019}, quantum radar\,\cite{CVB19, DCA19, RRM2020, PVS20, MR20}, and quantum  communication\,\cite{JHS09, ZTZ13, SZW14}.
	
	The standard approach for QI hinges on the use of entangled photon pairs as signal and idler probes for object detection\,\cite{SL08}. Here, the signal beam navigates through a region containing the target object amid background noise, while the idler beam is directly send to the receiver and retained  until the reflected signal from the object reaches the receiver.  What sets QI apart and enhances its performance over classical counterparts is its strategic application of detection and joint measurement techniques to calculate quantum correlations using the idler and signal beams. Only when
	a reflected signal from the object is present do these techniques efficiently capture nonclassical correlations between the stored idler and the reflected signal, affirming the object's presence while concurrently mitigating background noise and false signal. Few approaches for QI not exploiting entanglement, but instead the strong temporal correlation intrinsic to photon pair, idler and signal photon generation from spontaneous parametric down conversion process is used to suppress background noise are being reported \,\cite{liu2019,mrozwoski2024,blakey2022,frick2020}. In such approaches, background noise that does not coincide with the idler photons is suppressed, outperforming classical illumination schemes. However, noise that coincides with the idler photons contributes to false signals. When entanglement between the signal and idler photons, in addition to temporal correlation, is used for object detection, false coincidence detection of noise with idler photons is eliminated while taking the orthogonal basis measurements required for calculating entanglement, resulting in the elimination of any false signal detection. Recent theoretical studies has also highlighted the enhanced effectiveness of QI when employing hyper-entangled states of light for probing \,\cite{PSC21,KC22}.The core objective of all approaches for QI measurements is to minimize uncertainty in estimating unknown parameters by exploiting quantum correlations. With access to multiple ways of generating entangled states of light and measuring quantum correlations within them,the topic leaves open the exploration of various configurations with the potential to extend the principles of target detection accuracy, ranging sensitivity, and resilience against prevalent noise.

	In previously reported quantum illumination (QI) experiments and proposals for its realization, various quantum correlation measures have been employed and proposed to detect the presence of an object.\,\cite{GIE09,JSL09, ZMW15, GMT20, ZYW21, GM2021}. Among them,  utilizing photon number correlation, general Cauchy Schwartz value ($\epsilon$)  is one such example.  The general Cauchy Schwartz value is achieved up to $\epsilon \approx 10$, where the classical bound being 1. The result, at an object reflectivity $\eta=0.5$ immersed in noisy thermal background was reported where the non-classicality prevails up to signal to noise ratio (SNR) $\approx 0.5$\,\cite{LBD13, LBO14}.  A QI system is also documented to exhibit a 10-fold increase in SNR when compared to its classical counterpart, even in the absence of a simultaneous measurement on the signal and reference beams\,\cite{CVB19}. Another theoretical paper suggested quantum error probability as a measure,  and reported a 6 dB advantage with entangled pairs over a coherent state system\,\cite{TEG08GS}.   Experimental demonstration of QI using symmetric polarization-entangled photon state has also been recently reported and it  relayed on Bell state measurement to confirm the presence of object. It also shows  that it breaks the classical limit  for up to a $40\%$,  while approaching the quantum limit imposed by the Helstrom limit when $\eta \ge 0.4$\,\cite{LBD14, CWH76}.  
	
	Here we propose and experimentally demonstrate a QI  scheme using polarization-entangled  photons which uses Bell's inequality measurement, CHSH value  as the quantum correlation measure.  The CHSH value,  $S$ is used for identifying the presence of object and normalised CHSH value, $\bar{S}$  is used for  estimating  the reflectivity of the object after detecting its presence.  Unlike other quantum correlation measures used for QI,  the CHSH value also helps us to identify the range of values of  $S$ where quantum correlations are absent but can be marked as a  residual of quantum correlation obtained due to the entanglement in the probe state. The value of $\textrm{S}>2$  in the scheme indicates the quantum correlation, and $\sqrt{2} \le \textrm{S} \le 2$ signifies classical residual to the quantum correlations  which can be used for confirming the presence of object\,\cite{QOQI_1}.  These bounds are fixed by using the maximum  $S$ value that can be achieved with the same polarizing angles for a separable state, that is  $S = \sqrt{2}$.  Our experimental results shows quantum correlation,  $S > 2$ up to object reflectivity $\eta =0.1$ and residual of  quantum correlation, $\sqrt{2} \le \textrm{S} \le 2$  for $\eta = 0.05$. Even in the presence of noise with a signal-to-noise ratio (SNR) of 0.03 we have obtained  $S >  \sqrt{2}$,  when $\eta=0.7$. While the signature of quantum correlation cannot be established when $0 < S < \sqrt{2}$ to confirm the presence of an object, it still indicates the object's existence, albeit with a possibility of false positives. However, in this work we focus only on the detecting an object using quantum correlation without any ambiguity The object's range can be estimated by calculating the delay in arrival time of the signal photon correlated in time with the idler photons\cite{mrozwoski2024}. Although quantum correlation measurements typically involve measuring in different quantum state bases, single-shot real-time measurement can be achieved by dividing the signal and idler paths into four channels each and measuring all required basis states simultaneously using eight detectors. The details of this method are described in the experimental methods section.

	Apart from considering the reflectivity of the object and SNR, another crucial aspect of object detection using a quantum probe is the estimation and effect on purity of quantum probes due to atmospheric attenuation\,\cite{ZLU22,QZ22}.  The studies in this domain have been conducted using a NOON state as a probe, where the atmosphere affects the interference pattern, and the standard deviation of an operator $\delta A$ is considered as a measure to obtain a bound\,\cite{Att_3}. For our purpose, we have modeled  single-photon attenuation due to the atmosphere and report how the CHSH value behaves with distance. Such attenuation effects have been studied previously in the context of quantum key distribution (QKD)\,\cite{Att_1,Att_2}. Following a similar approach, we consider a constant atmospheric attenuation and investigate the single-photon attenuation due to the atmosphere. We report how the CHSH value varies with distance. Our estimations indicate that we can achieve $S > \sqrt{2}$ and identify the presence of an object at distances up to 25 km. This distance cover the lateral distance of  atmosphere where attenuation of photons happen in general.


	\section{Quantum illumination using polarization-entangled photon pairs } 
	\label{sec2} 
	
\begin{figure}[h!]
		\centering
		\includegraphics[width= 0.7\linewidth]{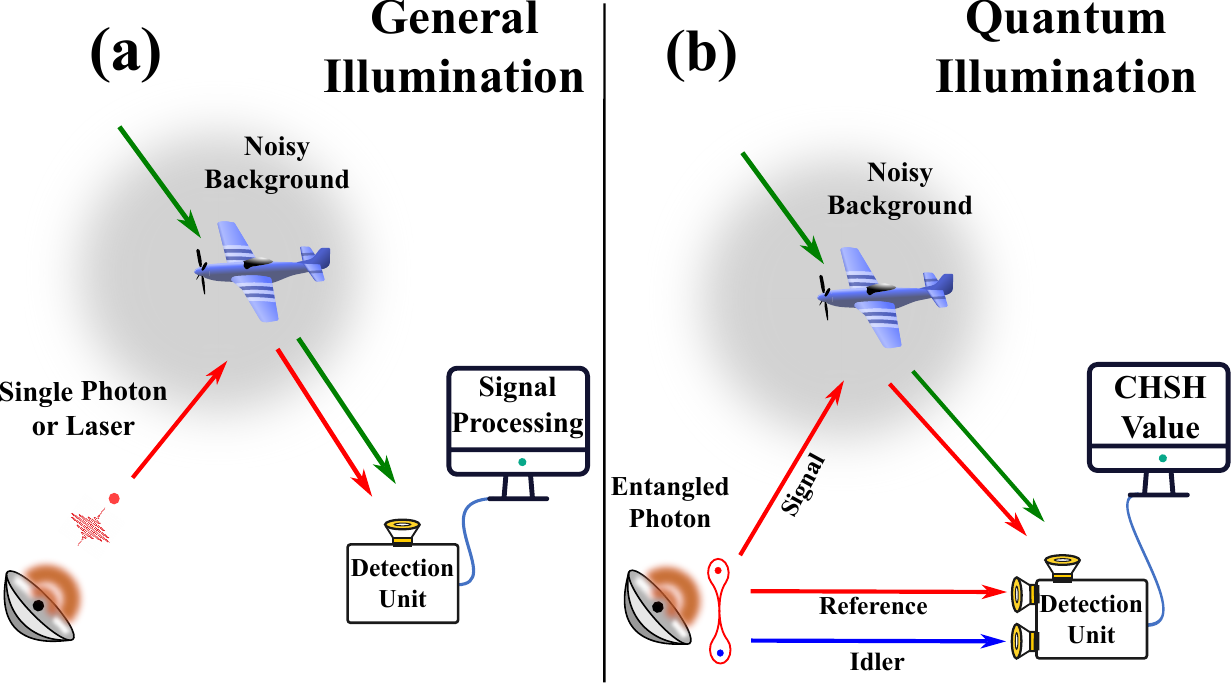}
		\caption{A schematic of the illumination protocol is depicted. (a) States, such as single photons or normal coherent states (laser beams), are employed as probe states for detecting objects in the presence of noise. Only direct measurements are included, with no joint measurements are involved. (b) Polarization-entangled photons are utilized. The signal beam is split into two; one beam is directed for object detection, and the other is sent to the measurement unit. Finally, measurements are conducted based on joint detection with the idler, and the CHSH values are calculated.}
		\label{Fig1}
	\end{figure}

	The illustration of the general illumination and our quantum illumination scheme using polarization-entangled photon pairs is shown in Fig.\,\ref{Fig1}. In general illumination, a single photon or coherent laser source is sent towards the object. Without a target, only noise is received. With a target present, a return signal accompanied by noise is detected. Isolating noise in both cases is challenging.  In the QI scheme, an entangled state comprises a signal photon directed towards the target and an idler photon sent directly to the receiver. The CHSH value between the returned signal photon reflected from the object and the idler photon at the receiver can isolate noise and confirm the object's presence.
	
	The polarization-entangled state of two photons from a spontaneous parametric down conversion (SPDC) process using  Beta Barium Borate (BBO) type-II  crystal is 
	\begin{equation}
		\ket{\Psi^{-}} = \frac{1}{\sqrt{2}}\left(\ket{H}_{s}\ket{V}_{i}-\ket{V}_{s}\ket{H}_{i}\right),
	\end{equation}
	where $|H\rangle$ and $|V\rangle$  represent the horizontal and vertical polarization states  of the photons, respectively, and subscripts $s$ and $i$ represent signal and idler photons, respectively.  To calculate the CHSH value and demonstrate Bell's violation $S \geq 2$, it is necessary to calculate the projection values at different polarizing angles in the signal and idler channels\,\cite{ENTG_1,ENTG_2}.
	
	The projection state for signal and idler photons at angle $\alpha$ and $\beta$ is  

		\begin{align}
			\ket{H_{\alpha}}_{s} &= \cos(\alpha)\ket{H}_{s} - \sin(\alpha)\ket{V}_{s} \notag \\
			\ket{V_{\alpha}}_{s} &= \sin(\alpha)\ket{H}_{s} + \cos(\alpha)\ket{V}_{s} \notag \\
			\ket{H_{\beta}}_{i} &= \cos(\beta)\ket{H}_{i} - \sin(\beta)\ket{V}_{i} \notag \\
			\ket{V_{\beta}}_{i} &= \sin(\beta)\ket{H}_{i} + \cos(\beta)\ket{V}_{i}
		\end{align}

	Based on these projection values, the probabilities of finding the photons in specific polarization combinations, denoted by the subscripts,  $P_{HH}$ , $P_{VV}$ , $P_{H,V}$ \& $P_{VH}$, are calculated for correlation parameter $E(\alpha,\beta)$ is expressed as  
	$$E(\alpha,\beta) = P_{HH} + P_{VV} - P_{HV} - P_{VH}$$ 
	which is then used to calculate the CHSH value, given by  
	\begin{equation}
		S = |E(\alpha,\beta)-E(\alpha,\beta^{'})+E(\alpha^{'},\beta)+E(\alpha^{'},\beta^{'})|.
	\end{equation}

	Experimentally, we calculate the $E(\alpha,\beta)$ using coincidence detection for various combination of the polarization angles,
	\begin{equation}
		E(\alpha,\beta) = \frac{[N(\alpha,\beta)+N(\alpha^{\perp},\beta^{\perp})-N(\alpha^{\perp},\beta)-N(\alpha,\beta^{\perp})]}{[N(\alpha,\beta)+N(\alpha^{\perp},\beta^{\perp})+N(\alpha^{\perp},\beta)+N(\alpha,\beta^{\perp})]}.
	\end{equation}
	
	Here $N(\alpha,\beta)$ represents the coincidence counts at the rotation angle of the polarization $\alpha$, $\beta$ in the signal and idler channels, respectively. Apart from the polarization if we also take into account the photon number in the both channels, then the complete quantum state becomes, \begin{equation}
		\ket{\Psi}_{\textrm{int.}} = \frac{1}{\sqrt{2}}(\ket{H}_{s}\ket{V}_{i}-\ket{V}_{s}\ket{H}_{i})\ket{1}_{s}\ket{1}_{i},
	\end{equation}
	which is a state vector  of the total  Hilbert space,  $\mathcal{H}_{\textrm{Pol}_{1}}\otimes\mathcal{H}_{\textrm{Pol}_{2}}\otimes\mathcal{H}_{\textrm{N}_{1}}\otimes\mathcal{H}_{\textrm{N}_{2}}$. 
	The $N_{1}$ and $N_{2}$ represent the number of photons in the signal and idler channels. To realize QI  we have to introduce an object in the channel of signal with a specific object reflectivity, $\eta$ and then measure the quantum correlation between the reflected photons from the object and the idler retained as our reference.
	
	We can realize the object with varying beam splitter reflectance ($\eta$) and consider photon loss in the signal channel as a Kraus operator ($K$) \cite{Kra_1,Kra_2}.  There are two approaches to model this situation. The first is to track only the photons that ultimately coincide after being reflected from the object, as our measurements focus solely on the coincidence counts of the photons. This approach is equivalent to modeling a photon loss channel, where we consider only the photons that are successfully detected, then the Kraus's operators become  
	
	\begin{equation}
		\label{K1}
		K_{0}^{s} = \sqrt{(1-\eta)}\ket{0}_{s}\bra{1}_{s} \, ; \, K_{1}^{s} = \ket{0}_{s}\bra{0}_{s} + \sqrt{\eta}\ket{1}_{s}\bra{1}_{s}
	\end{equation}
	
	which  satisfy 
	$$\sum_{j=0}^{1}(K_{j}^{s}(\eta))^{\dagger}K_{j}^{s}(\eta) = \mathbb{I}.$$
	For the idler photon part, there is no photon loss, so it remains as $K^{i}=\ket{1}_{i}\bra{1}_{i}$. The second approach is not to keep track of the lost photons. In this case, the photon-loss channel, Kraus operator takes the form 
	\begin{equation}
		\label{K2}
		M^{s}_{1} = (1-\eta)\ket{0}_{s}\bra{0}_{s} + \eta\ket{1}_{s}\bra{1}_{s}. 
	\end{equation}
	If $\eta > 0$, then  $M^{s}_{1}$ operator satisfy  $$(M_{1}^{s}(\eta))^{\dagger}M_{1}^{s}(\eta) < \mathbb{I}.$$
	Finally, the total Kraus operator becomes, 
	$$ M = M_{1}^{s}\otimes M_{1}^{i} = ((1-\eta)\ket{0}_{s}\bra{0}_{s}+ \eta\ket{1}_{s}\bra{1}_{s})\otimes \ket{1}_{i}\bra{1}_{i}.$$ 
	The density matrix of the initial state ,  $\rho_{int.} = \ket{\Psi}_{\textrm{int.}}\bra{\Psi}_{\textrm{int.}}$.
	The density matrix considering to photon loss channel becomes,  
	\begin{equation}
		\rho(\eta) = (\mathbb{I}_{2}\otimes \mathbb{I}_{2} \otimes M)\rho_{\textrm{int.}} (\mathbb{I}_{2} \otimes \mathbb{I}_{2} \otimes M)^{\dagger}.
	\end{equation}
	The operation $\mathbb{I}_{2}\otimes\mathbb{I}_{2}$  indicates that the photon loss does not change the polarization information. Finally, to calculate the quantum correlation between the reflected photon and the reference photons, we trace out the photons number information and then check the CHSH value of the reduced density matrix,
	\begin{equation}
		\rho_{\textrm{red}} =  \textrm{Tr}_{N_{1},N_{2}}(\rho(\eta)).
	\end{equation}
	In the first case discussed Eq.\,(\ref{K1}), the CHSH value remains constant and maximum even when the object's reflectivity is very small. This helps us to confirm the presence of the object but does not assist in estimating the value of $\eta$ for the object. However, in the second case Eq.\,(\ref{K2}), as the photon loss in the signal path increases, we record a decrease in the CHSH value, which helps estimate the reflectivity of the object. We refer to this as the normalized CHSH value, $\bar{S}$. 
	
\begin{figure}[h!]
		\centering
		\includegraphics[width= 1.0\linewidth]{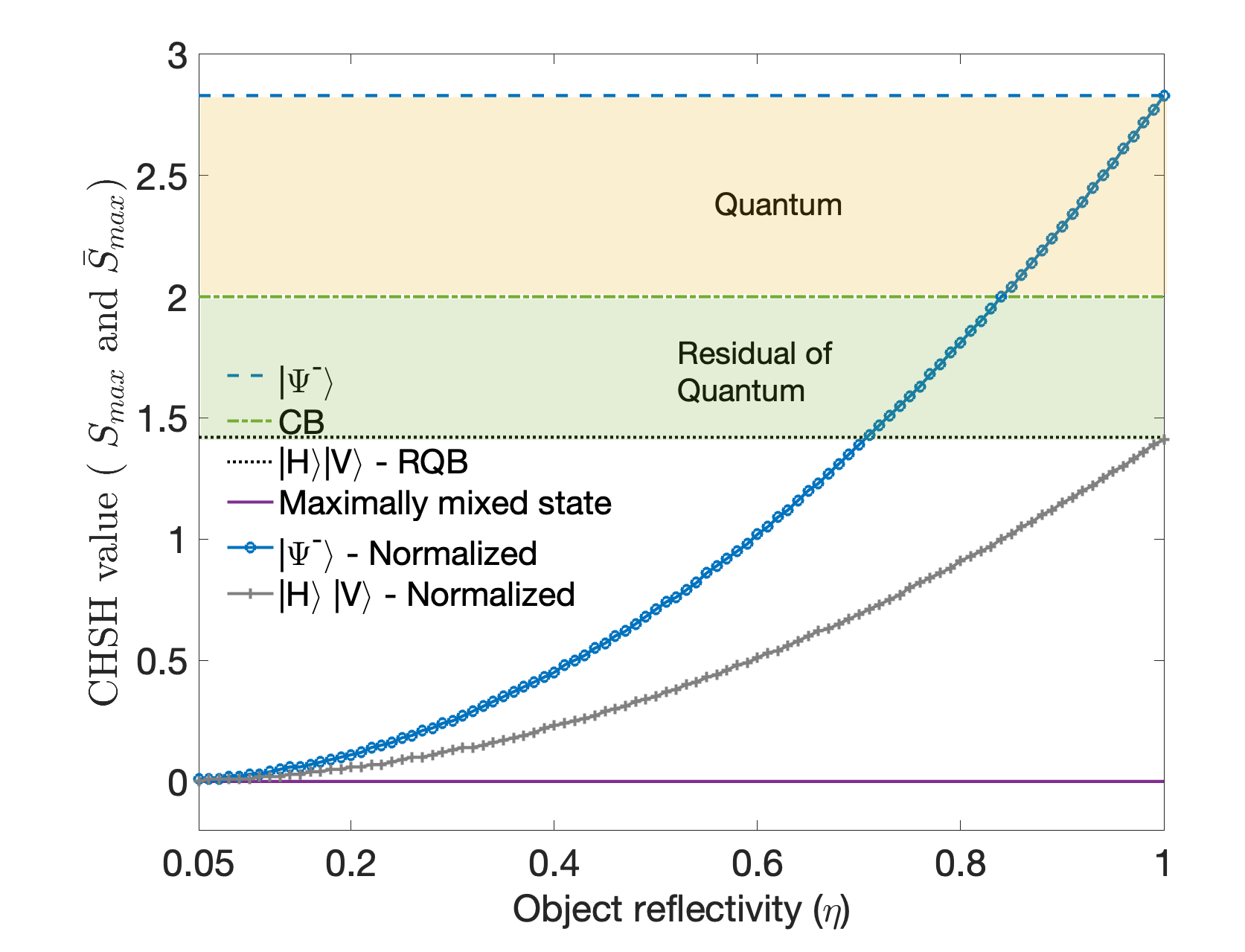}
		\caption{Theoretically calculated CHSH values using different, entangled,  pure separable, and maximally mixed state as probes  for two different theoretical models are presented as a function of object reflectivity ($\eta$) ranging from  0.05 to 1. Depending on the maximum value we can obtain using each probe state, the quantum bound, classical bound (CB) and the residual quantum bound (RQB). The region within the bounds are shown in different colors. For all the probe states the standard CHSH value ($S_{max}$) shows different value that sets the bounds but remains constant for all non-zero $\eta$.  Except for maximally mixed state, normalised CHSH value ($\bar{S}_{max}$) decreases with decrease in $\eta$. For maximally  mixed state in both scenario, CHSH-value remain 0.}
		\label{Fig2}
	\end{figure}

	In Fig.\,\ref{Fig2}, we present the CHSH value and the normalized CHSH value, theoretically calculated as functions of object reflectivity $\eta$, using different states of light as probes for illuminating the object.  We can see that the CHSH value remains constant for all $\eta$ with maximum value. The value of $S_{max} = 2\sqrt{2}$ for the maximally entangled state  and minimum value $S_{max} = 0$ for the maximally mixed state.  When unentangled pure state of photons pairs, $|H\rangle_{s}|V\rangle_{i}$ are used as signal and idler photons, the $S_{max}$ becomes $\sqrt{2}$. Therefore, the CHSH value,  $2\sqrt{2} \geq S \geq \sqrt{2}$ can be used as a range of values for confirming the presence of an object.  The value of $S>2$ indicates the presence of quantum correlation and value, $2 \geq S \geq \sqrt{2}$  indicate classical correlation which can be called as residual of quantum correlation.  Since these values cannot be achieved with unentangled states as probes, they indicate entangled probe states in a noisy environment and are referred to as residual quantum correlations. The normalized CHSH value for both an entangled state and pure unentangled two-photon states shows a decrease with a decrease in $\eta$.  Therefore, using the CHSH value, $S \geq \sqrt{2}$ and the normalized CHSH value (even when less than $\sqrt{2}$), we can confirm the presence of the object and quantify the object's reflectivity.
	
	In our experimental setup with entangled photon pairs, the determination of the CHSH value involves analyzing coincidence counts. While experimental calculations yield various object reflectivities \(\eta\), the CHSH value (\(S\)) remains constant, signifying that the reflected photons retain entanglement despite photon loss. This aligns with the first scenario mentioned earlier.
	
	To illustrate the second scenario, we introduce a normalization of the correlation value \(E(\alpha,\beta)\)  with respect to our reference denoted by $E^{'}(\alpha,\beta)$. Specifically,
	
	\begin{equation}
		E^{'}(\alpha,\beta) = \frac{[N(\alpha,\beta)+N(\alpha^{\perp},\beta^{\perp})-N(\alpha^{\perp},\beta)-N(\alpha,\beta^{\perp})]_{\text{Act}}}{[N(\alpha,\beta)+N(\alpha^{\perp},\beta^{\perp})+N(\alpha^{\perp},\beta)+N(\alpha,\beta^{\perp})]_{\text{Ref}}}.
	\end{equation}
	
	Here, $S$ value ($\Bar{S}$) obtained form $E^{'}(\alpha,\beta)$  differs from the CHSH value but matches the theoretical plot obtained for to the second Kraus operator. This normalization allows us to extract information about the object's reflectivity.
	
	In the presence of both object-induced losses and depolarizing noise in the object channel, the photon's polarization information is directly affected. The depolarizing noise, characterized by a single parameter \(p\), is represented by Kraus operators,	
	\[
	K_{0} = \sqrt{1-\frac{3p}{4}}I, \; K_{1} = \sqrt{\frac{p}{4}}X, \; K_{2} = \sqrt{\frac{p}{4}}Y, \; K_{3} = \sqrt{\frac{p}{4}}Z.
	\]
	Since we deal only with the depolarization of the photon, this set of Kraus operators satisfy the condition $\sum_{i} K_{i}^{\dagger}K_{i} = \mathbb{I}$,
	The density matrix for the depolarizing noise alone is given by:
	\begin{equation}
		\rho(p) = \sum_{i=0}^{3}(K(p)_{i})^{\dagger}\rho (K(p)_{i}).
	\end{equation}
	
	For a realistic situation involving both the presence of an object and depolarizing noise, the evolved density matrix can be expressed by two parameters: the object reflectivity \(\eta\) and depolarizing parameter $p$,
	\begin{equation}
		\rho(p,\eta) = \sum_{i=0}^{3}(K(p)_{i})^{\dagger}\rho(\eta) (K(p)_{i}).
	\end{equation}

	\section{Experimental Method }

	\subsection{Experimental Setup}
	\begin{figure*}[t!]
		\centering
		\includegraphics[width= 1\linewidth]{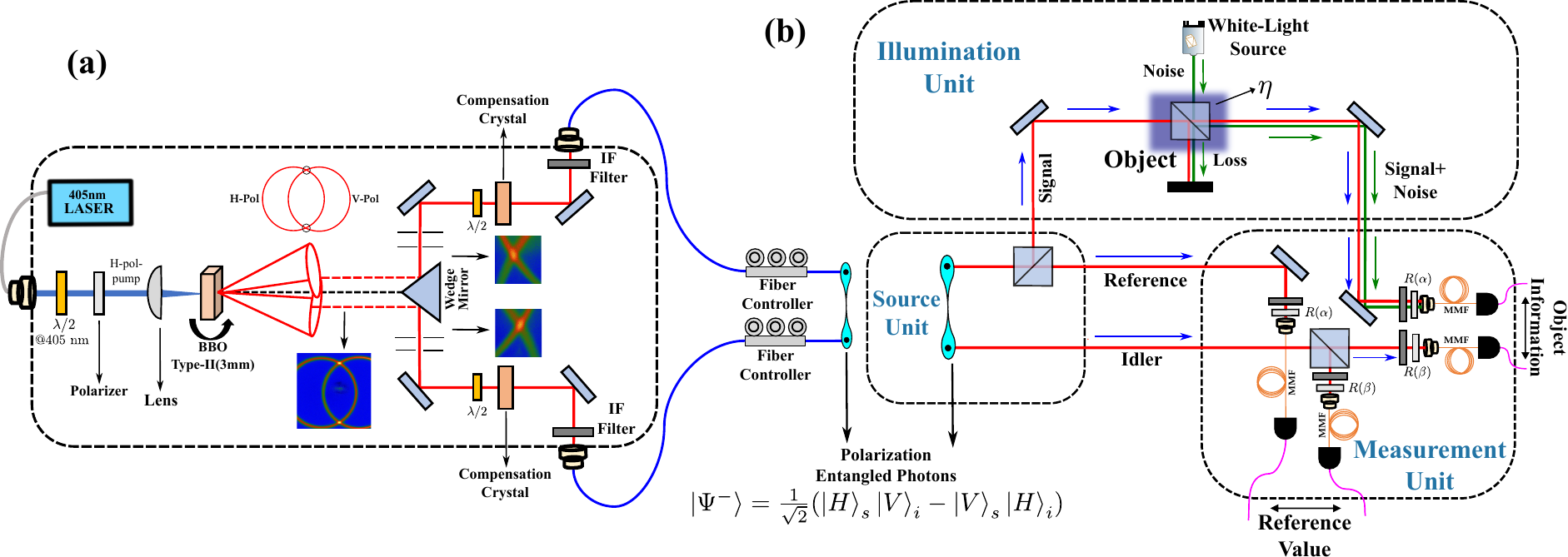}
		\caption{Experimental setup: (a) This section describes the preparation of entangled photons using a type-II BBO crystal. The entangled pairs, generated through SPDC, are separated via a wedge prism and coupled with a fiber. Fiber controllers are employed to achieve the desired polarization, ensuring that the entanglement persists after passing through the fiber.(b) The illumination setup is divided into three sections: (i) Source Unit (SU), (ii) Illumination Unit (IU), and (iii) Measurement Unit (MU). In SU, the signal-photon beam is divided into two channels: one directly goes to the MU, and the other goes for object detection. Idler photon beam, which serve as a reference, directly go to the MU and there its split into two. The CHSH value between the two signal channel, one which is sent to the object and other that directly goes to the MU, is calculated using the idler pair. The object is placed on a beam-splitter with different object reflectivity, and white-halogen light serves as noise introduced in the object (BS) arm.}
		\label{Fig3}
	\end{figure*}
	Polarization-entangled photons at 810 nm are generated from an  SPDC process  using a BBO nonlinear crystal. The crystal is type-II phase matched with a thickness of 3 mm. The schematic of the experimental setup for generating entangled photons pairs is shown in Fig. \ref{Fig3}(a). The BBO crystal is pumped using a continuous-wave diode laser (MatchBox, Integrated Optics) at 405 nm with a spectral line width of 0.6 nm and a pump power of 10 mW. A half-waveplate at 405 nm is used to set the pump polarization perpendicular to the optical axis of the BBO crystal. An achromatic lens of 25 mm is used to tightly focus the pump laser into the centre of the crystal. The spatial distribution of the horizontal and vertically polarized SPDC photon pairs is captured using an EMCCD camera (Andor iXon). A typical image of the generated SPDC photon pairs is shown in the inset of Fig.\,\ref{Fig3}(a). The two intersection regions in the image are the polarization-entangled photons.
	
	A wedge-mirror is used to direct the entangled photons into two arms for ease of coupling to a single-mode optical fiber. To compensate for the transverse and longitudinal walk-off of horizontally polarized photons to the vertically polarized photons, we employ a half-waveplate at an angle of $\pi/4$ and BBO compensation crystal of 1.5 mm thickness in both arms. The residual pump photons are filtered using an interference filter ($810\pm10$ nm) before entering to the single-mode fiber. The single-mode fibers are preloaded with a fiber polarization controller to maintain the actual polarization of the photons.
	
	To quantify the polarization entanglement, the fiber-coupled entangled photons are sent to fiber-coupled single-photon counting modules (SPCMs) (Excelitas) after passing through a polarizer. The output of the SPCMs are fed in to a time-correlated single photon counter (Time Tagger, Swabian Instruments). Initially, the coincidence visibility of the photons in the horizontal (H)- vertical (V) and Anti-diagonal (A)- diagonal (D) polarization basis are maximized. The visibility in the H/V and A/D basis is maximized by tilting the BBO compensator crystals and adjusting fiber paddles, respectively. The measured visibility of the polarization entangled source in the H/V basis is around 97\% while in the A/D basis, it is around 94\%. Fig.\,\ref{fig:P6} in the Appendix shows a typical coincidence visibility curve for all the four bases. The calculated CHSH value for the source is $2.72\pm0.05$.

	Figure \ref{Fig3}b shows the schematic of the experimental setup used for quantum illumination with polarization-entangled photon pairs. The generated fiber-coupled signal and idler photon counts are 7000 counts per second (c/s).
	
	In the entangled photon pair, the signal beam is split into two by a 50:50 beam splitter. One arm is used to illuminate the object, while the other part goes to the measurement unit (reference arm in Fig.\,\ref{Fig3}). The idler arm, however, goes directly to the measurement unit, where it is split into two using a 50:50 beam splitter. We then calculate the correlation between one of the split idler arms and the signal's reference arm, and between the other split idler arm and the photons reflected from the object. In the measurement unit, $R(\alpha)$ and $R(\beta)$ represent the specific polarizer angles used to calculate the correlation parameter $E(\alpha,\beta)$ using the coincidence counts, leading to the final S-parameter.
	
	In comparison to the other QI schemes, if we divide the signal and idler paths into two channels each, and then use a half-wave plate (HWP) and a polarizing beam splitter (PBS) in each path, we can adjust the HWPs to appropriate angles which maximize correlation parameter $E(\alpha,\beta)$. After the PBS in each arm, a total of 8 paths are created, requiring 8 detectors. Finally, we need to check the coincidences between the four signal arms and the four idler arms, considering all possible combinations simultaneously. This implies we need to measure the coincidences between each signal arm and all the corresponding idler arms at the same time. As a result, we can perform single-shot measurements in real time. The process of measuring coincidences across all the arms and calculating the correlation parameter depends on the detector dead time, time tagger resolution and the brightness of the entangled photon source. With higher photon counts, we can reduce the coincidence integration time, which helps to reduce the overall duration of the measurement significantly.
	
	The object used in the experiment are beam splitters with varying reflectivity, $\eta$, it is varied from 1 to 0.05. Finally, we calculated the CHSH value between the photons reflected from the object and the transmitted arm of the idler. This procedure allows us to analyze the object's reflectivity in real time, as illustrated in Fig.\,\ref{Fig3}(c). The motivation behind referencing the signal and idler is to calculate quantum correlations for two types of scenarios using two different sets of Kraus operators, is discussed in the theoretical description.

	However, while demonstrating the effects of noise and depolarizing effects, we primarily relied on the direct measurement of the CHSH value from the object arm and one of the idler arms. In this case, we deal only with the first type of Kraus operator.
	
	To investigate the impact of noise, we have incorporated a fiber-coupled white halogen light source (Ocean-Insight. HALOGEN, HL-2000-FHSA) and connected it to one of the arms of the object beam splitter, as illustrated in Fig.\,\ref{Fig3}(b). This is a tunable thermal light source, allowing us to adjust it to attain various SNR  values. We varied the SNR from 1 to $2\times10^{-3}$. 
	
	Realizing depolarizing noise is experimentally challenging. One approach to address this involves using two calcite crystals and a quarter-wave plate in the signal arm, with the rotation angle of the quarter-wave plate serving as the depolarizing value\,\cite{De_pol_1, De_pol_2}. Another approach we follow is to introduce a combination of a quarter-wave plate, a half-wave plate, and another quarter-wave plate (Q-H-Q) in the signal arm\,\cite{QHQ_1, QHQ_2}. The quarter-wave plates are fixed at angles of $-\pi/4$ and $\pi/4$, while the half-wave plate is rotated to serve as the depolarizing element. Then the quantum state becomes, 
	
	\begin{equation}
		\ket{\Psi}_{\textrm{De-pol}} = -ie^{-i2\theta}\big\{ \ket{\Phi^{+}}+\left(\frac{1-e^{-4i\theta}} {\sqrt{2}}\right)\ket{V}\ket{V}\big\}
		\label{Eq7}.
	\end{equation}
	
	The second  term in the Eq.\,(\ref{Eq7}) equation reflects the depolarizing factor. As $\theta$ increases the depolarizing effect increases, and reaches its maximum when the HWP angle is $\pi/4$. This is because in half-waveplate we have a $2\theta$ dependence, and in expression, we got $4\theta$ term, so depolarizing value $p$ becomes 1 when the half-waveplate angle becomes $\pi/4$. Although all Q-H-Q operations are unitary, ideally they do not change the entanglement at all. However, due to the specific measurement setup we use, the S-value drops for the transformed state described in Eq.\,(\ref{Eq7}). This setup is used solely to model the depolarizing situation.

	\subsection{Experimental result and analysis} 
	
	\begin{figure}[h!]
		\centering
		\includegraphics[width= 1.0\linewidth]{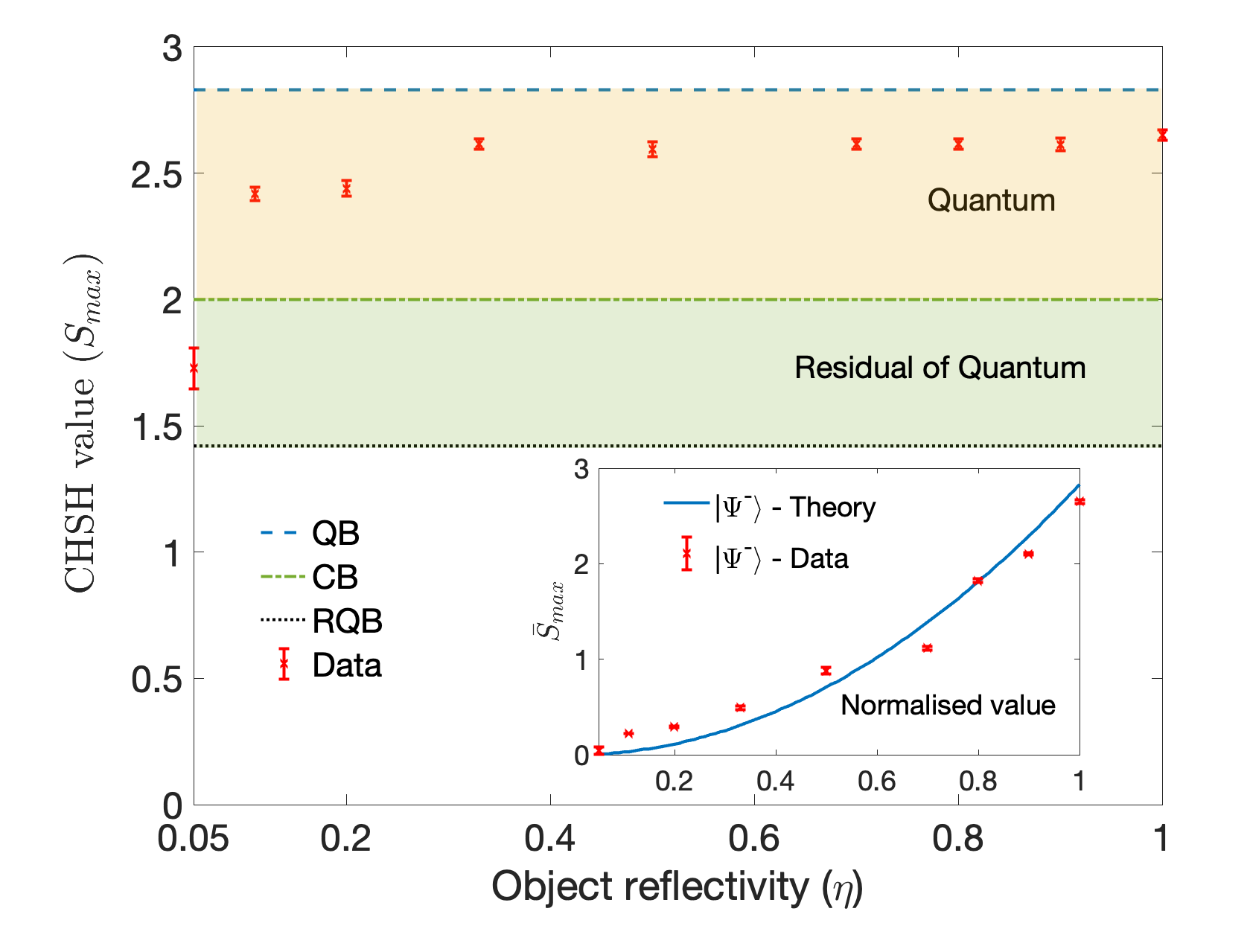}
		\caption{Experimentally obtained maximum value of CHSH value, ${S}_{max}$ for an object of different reflectivity($\eta$). In the main figure the red-points shows the experimentally obtained data for various object reflectivity, and at $\eta \approx 0.05$ the CHSH-value goes below 2, but lies in the residual of quantum region. Decrease in $S$ indicate the limit of  reflected photons getting close to the dark counts of the detectors. The-inset figure shows how the normalised CHSH-value($\bar{S}_{max}$) changes with different $\eta$, where the blue curve represents the theoretical modeling.}
		\label{Fig4}
	\end{figure}

	Figure \ref{Fig4} shows the experimentally obtained CHSH-value,  $S_{max}$ when objects of different reflectivity $\eta$ were illuminated using polarization-entangled photons. The upper shaded region in the plot shows the bound on the value of $S_{max}$ to be quantum. The data points with error bars show the value of $S_{max}$ for different object reflectivity. It should be noted that the $S_{max}$ remains the same for object reflectivity, $\eta$ up to 0.3. However, when the object reflectivity is 0.2, the reflected signal from the object is 700 c/s and when the object reflectivity is  0.05, the reflected signal from the object further decreases to 175 c/s, which is comparable to the dark count of the detector. 
	As a result of the random effect of dark counts, the $S$-value dropped below 2 for $\eta=0.05$, but it remains in the residual quantum region. A single measurement involves calculating the correlation parameter \( E(\alpha,\beta) \) for 4 different sets of angles (resulting in 16 combination coincidence detection values) to determine the S-parameter. Each data point in all the experimental plots is the avarage of 3 measurements. We set the integration time for the coincidence measurement as 5 seconds due to the low brightness of the source. The inset in Fig.\ref{Fig4} shows the effect of object reflectivity on $\bar{S}_{max}$. The data points with error bars represent the experimental data, while the theoretical model(solid line) is shown for comparison. Thus, experimental results confirm that using $S_{max}$ we can confirm the presence of the object and by using $\bar{S}_{max}$ we can estimate $\eta$.

	\begin{figure}[h!]
		\centering
		\includegraphics[width = 0.5\textwidth]{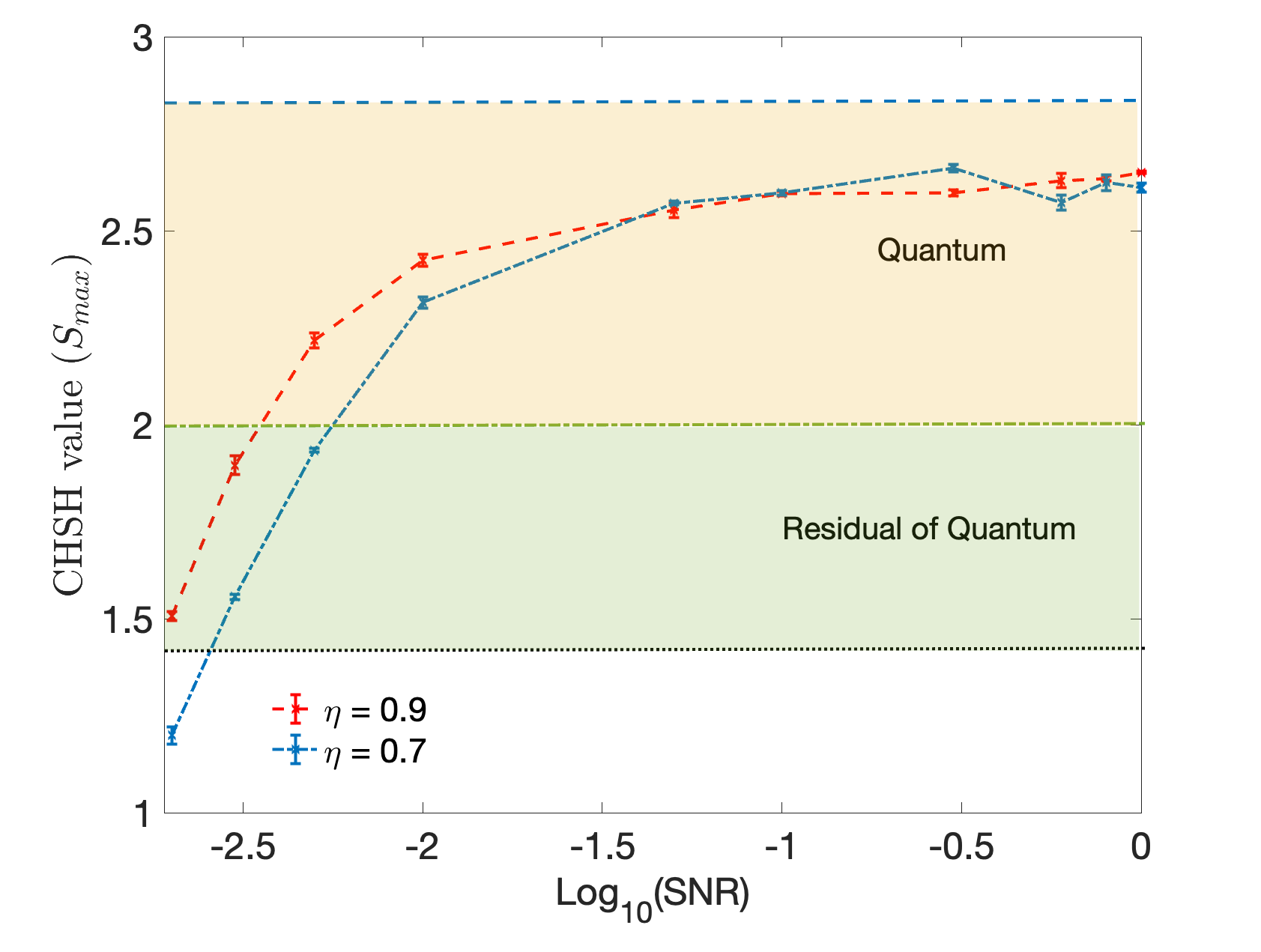}
		\caption{Experimentally obtained maximum of CHSH value for different SNR. Note that the x-axis is log of SNR.  The CHSH value remains within the quantum and residual of quantum bound for SNR value as low as 0.002 for $\eta =0.9$ and for SNR value of 0.03 for $\eta = 0.7$. Here the CHSH-value drops due to false coincidence counts.}
		\label{Fig5}
	\end{figure}
	
	Figure \ref{Fig5}, shows the experimentally obtained maximum of CHSH value when thermal noise is introduced. The red data points (square) and blue data points (asterisk)  shown are  for object reflectivity, $\eta$ of 0.9 and 0.7, respectively. By keeping the signal counts fixed, we increased the thermal noise. The SNR was varied from 1 to $2\times10^{-3}$. In Fig.\ref{Fig5}, the SNR is  plotted in $\log_{10}$ scale. One can see that with increasing the thermal noise the $\textrm{S}_{\textrm{max}}$ value remains almost the same for  SNR above 0.3 ( -1.522 ).

	For object reflectivity  $\eta= 0.9$ further increasing the noise level results in slow decrease in $S_{max}$ from 2.598$\pm$ 0.008 to 2.217$\pm$0.019 when SNR reaches  0.005 (-2.30)]. Further increasing the noise level, the $S_{max}$ drops below to 2 but still inside the residual of quantum bound, $S_{max}=  1.507\pm0.012$ at SNR of 0.002 (-2.69). For object with reflectivity $\eta = 0.7$,  a similar trend is seen. However at SNR 0.002 (-2.69), the $S_{max}$ value becomes 1.199 $\pm$0.007. These results indicate the decrease is $S$ value is seen only when the reflected photon detection counts reaches the limits of detector dark counts. 
	
\begin{figure}[h!]
		\centering
		\includegraphics[width=0.5\textwidth]{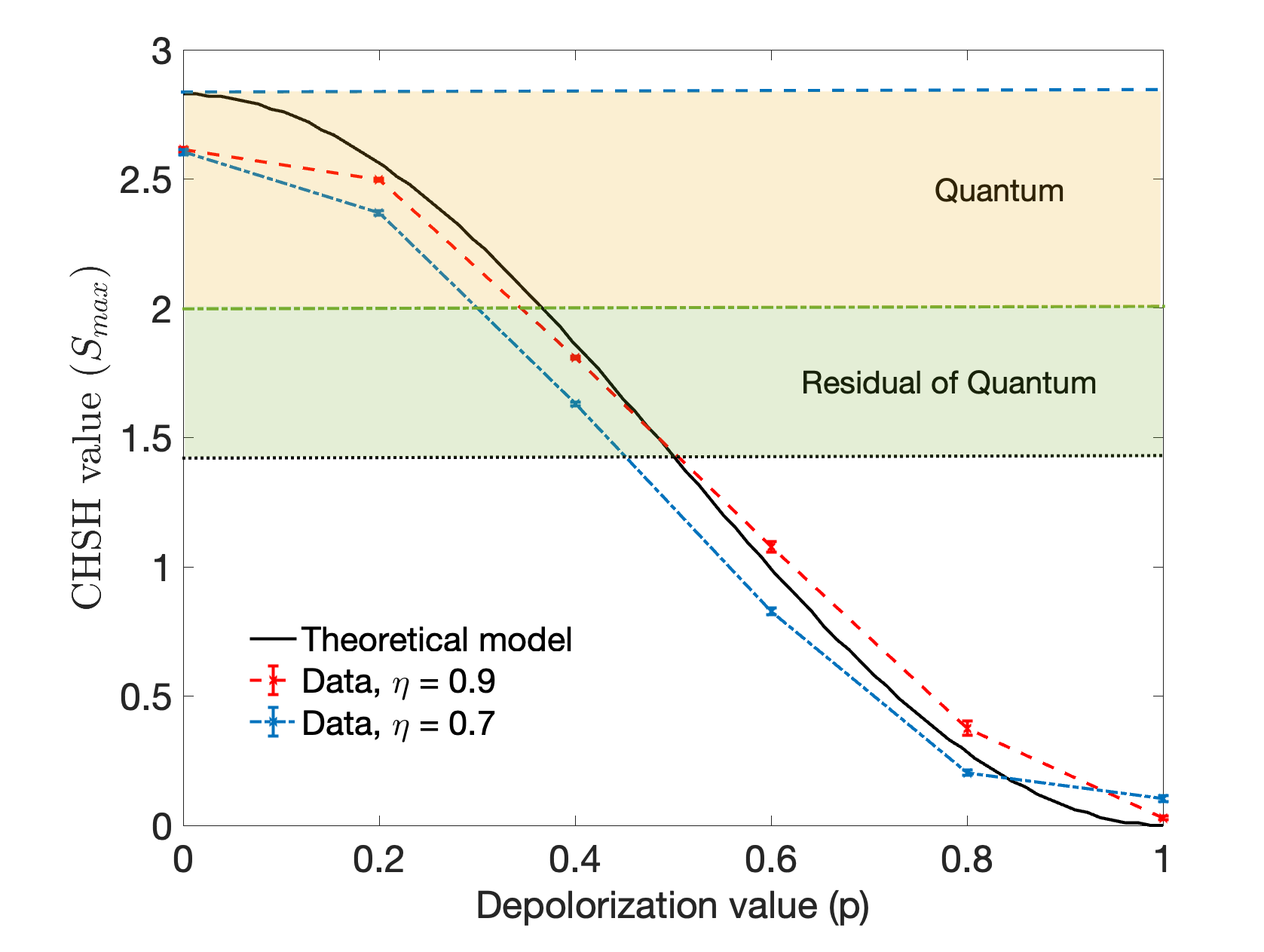}
		\caption{ Maximum  of CHSH value, ${S}_{max}$. The red data points (square) and blue data points (asterisk) show the experimentally obtained ${S}_{max}$ for $\eta$, 0.9 and 0.7.  with a change in depolarization value. The solid line shows the theoretical model using  Eq. \ref{Eq7}.}
		\label{Fig6}
	\end{figure}

	Figure \ref{Fig6} shows the maximum value of $S_{max}$ as a function of depolarization value for object reflectivity, $\eta =$ 0.9 and 0.7. The depolarization value was estimated from the polarization visibility of entangled photons by varying the half-waveplate in the Q-H-Q as mentioned in the second term of Eq.\,\ref{Eq7}. The red data points (square) and blue data
	points (asterisk) show the $S_{max}$ for $\eta = 0.9$ and $\eta=0.7$. The solid line shows the theoretical model using  Eq.\,\ref{Eq7}.  The observed data matches with the theoretical model.

	From these results we can say that the scheme is highly robust against thermal noise and can still be effective for against depolarization noise level of up to $p=0.5$.

	\subsection{Attenuation and range estimation} 
	
	To study the atmospheric effects, we need to model photon attenuation in the signal channel. In general, for a quantum state consisting of an N-photon Fock state $\ket{N}\bra{N}$, attenuation is modelled by introducing a distance variable ($L$). The number of photons that pass through the medium can be described by a binomial distribution, resulting in the final Fock state:
	\begin{equation}
		\rho(L) = \sum_{j=0}^{N} {N \choose j} (e^{-\Lambda L})^{j} (1 - e^{-\Lambda L})^{N-j} \ket{j}\bra{j}.
	\end{equation}
	Here, $\Lambda$ is the attenuation coefficient defined as $\Lambda \equiv \frac{\alpha \ln(10)}{10}$, and $\alpha = 0.07 \ \text{dB/km}$ for free space at a distance of $2.4 \ \text{kms}$. For our calculations, we consider $\alpha$ as a constant because, it decreases with increasing altitude, as the atmosphere starts to diminish at higher altitudes, and is slightly higher at lower altitudes. Therefore, for our purposes, we treat it as a constant.
	\begin{figure}[h!]
		\centering
		\includegraphics[width=0.5\textwidth]{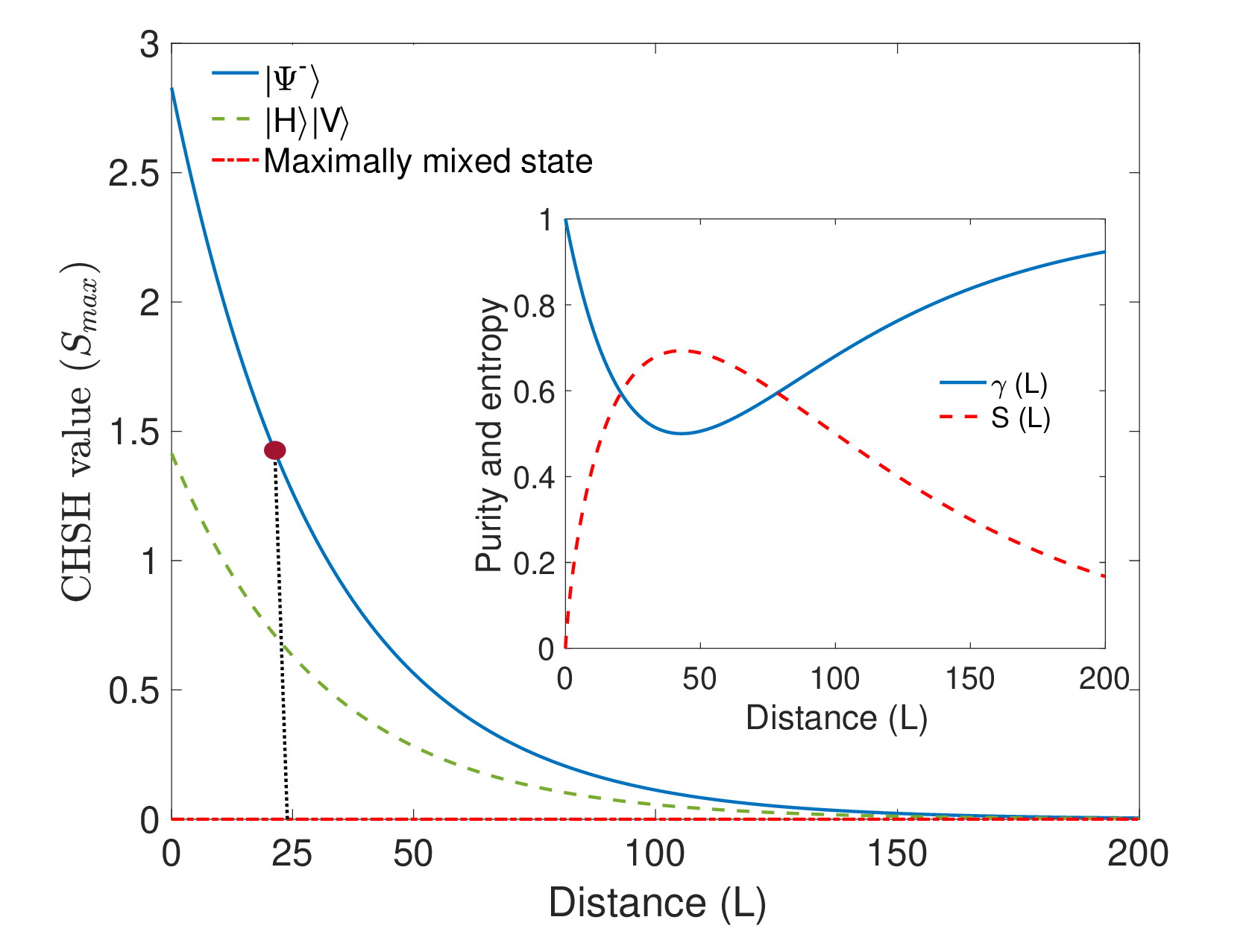}
		\caption{Theoretically, the change of CHSH value for the atmospheric attenuation of a single photon is shown as a function of distance (in kilometers) in the main figure, where the blue curve represents the entangled state $\ket{\Psi^{-}}$, the green dotted line corresponds to the state $\ket{H}\ket{V}$, and the red line represents the totally mixed state. From the figure, it is evident that the lower bound of classical residual of quantum persistence is approximately at $25$ kms. The inset figure shows the purity ($\gamma$) and entropy ($S$) of the single photon Fock state as a function of distance.}
		\label{fig:P5}
	\end{figure}

	Finally, we study the purity of the quantum state and Von-Neumann entropy,
	\begin{equation}
		\gamma(L) := \textrm{Tr}(\rho^{2}(L)) \,\, ; \,\,  S(L) :=- \textrm{Tr}\{\rho(L)\ln(\rho(L))\}.
	\end{equation}
	For our quantum illumination model, we deal with $\ket{0}$ and $\ket{1}$ only, then attenuation due to atmospheric effect the Fock state of the signal would become,
	\begin{equation}
		\rho_{s}(L) = (1-e^{-\Lambda L})\ket{0}_{s}\bra{0}_{s}+e^{-\Lambda L}\ket{1}_{s}\bra{1}_{s}.
	\end{equation}
	We can also check how the CHSH value changes with this distance to get an idea up to what range our quantum illumination protocol is valid.

	In Fig.\,(\ref{fig:P5}), the CHSH value with distance in the presence of atmospheric attenuation of photons is shown. We can observe that, for distances up to 25 kilometers, residual quantum correlations can be recorded. This distance falls within the range of two-way propagation of photons in the atmosphere, vertically upwards, up to the height where attenuation occurs.

	\section{Conclusion}

	In summary, we have experimentally demonstrated QI  protocol using polarization-entangled photon pairs based on the CHSH value as a measure of quantum correlation.  To begin with  we proposed the theoretical  model for the scheme, and demonstrated that our experimental results aligns well with the theoretical predictions. We calculated quantum correlation and normalised value of the quantum correlation across various object reflectivities to demonstrate the detection of object  and its reflectivity, respectively.  We have  assessed the impact of noise on quantum correlation, and shown the transition of correlation value for quantum to classical.  In addition, our study  quantify the range of CHSH value which captures only classical correlation but can be called and used as the residual of quantum correlation for illumination purpose.  As those value are obtained only when we have entanglement in source, they qualify to be  used for illumination and extend the limit of confirming the presence of  low reflectivity object even in absence of quantum correlation.   Moreover,  this paper shows the explicit use of residual of quantum correlation identify the presence of an object of low reflectivity even when SNR is as low as 0.003.  To extend the scope and robustness of the scheme,  we examined the effect of depolarizing noise in the signal arm and its influence on the CHSH value. Furthermore, we theoretically modeled a photon attenuation scheme to determine the distance over which our scheme maintains quantum correlations.  In summary,  this paper shows the explicit use of  CHSH value as  quantum ad residual of  quantum correlation to identify the presence of an object of low reflectivity even when SNR is as low as 0.003. 

	\section{Appendix}
	\begin{figure}[h!]
		\centering
		\includegraphics[width=0.5\textwidth]{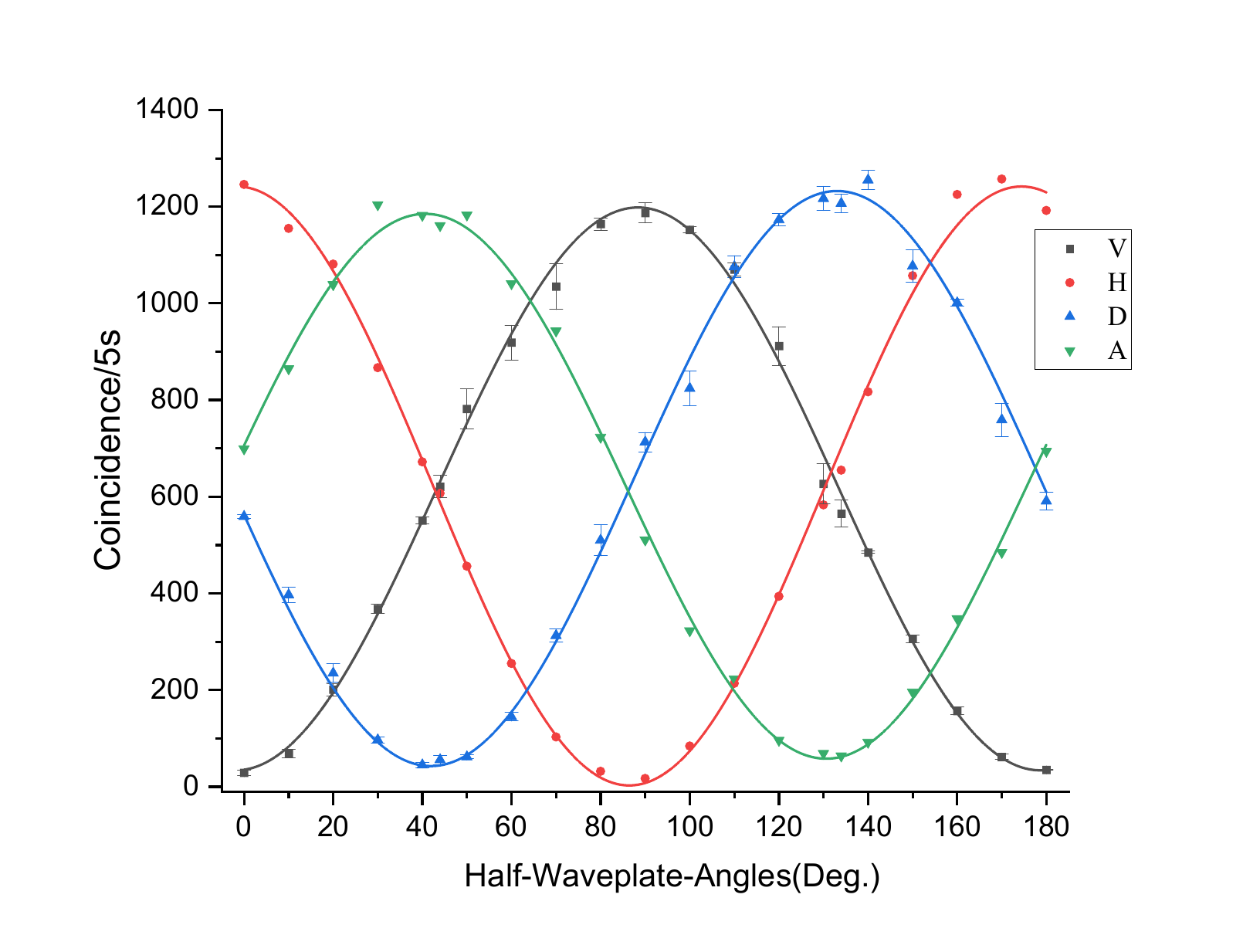}
		\caption{Experimentally, we obtained the visibility pattern of the polarization-entangled photon pair after passing through the fiber controller. The measured visibility in the H/V basis (red circle/black box) is around 97\%, while in the A/D basis (green down triangle, blue up triangle), it is around 94\%}
		\label{fig:P6}
	\end{figure}
	The visibility pattern of the entangled photon, for different basis after the fiber controller. To obtained the visibility pattern we employed a half-waveplate and PBS which act as an analyzer. 
	

	\section{Acknowledgment}
		We acknowledge the financial support from the Office of Principal Scientific Advisor to Government of India, project no. Prn.SA/QSim/2020.

		KS would like to thank Mayank Joshi and Mayalaxmi K for useful discussions on theoretical aspects.

		All data needed to evaluate the conclusions in the paper are present in the paper.  Data used for plots presented in the paper are available from the  corresponding author upon request.

		


\end{document}